\documentclass[reprint,lengthcheck,superscriptaddress,floatfix,nofootinbib,preprintnumber,longbibliography]{revtex4-1}

\usepackage{graphicx} 
\usepackage{amsmath,amsfonts,mathrsfs,amssymb}
\usepackage{microtype}
\usepackage{bbm}
\usepackage[toc,page]{appendix}
\usepackage{subfigure}
\usepackage{braket}
\usepackage{stackengine}
\usepackage{array,float}
\usepackage{cancel}
\usepackage{mathtools}
\usepackage{indentfirst}
\usepackage{adjustbox}
\usepackage{enumitem}
\usepackage{xcolor}

\usepackage[colorlinks=false,citecolor=blue,linkcolor=black]{hyperref}
\definecolor{JM}{rgb}{1.0, 0.31, 0.0}
\begin{document}
\preprint{LA-UR-20-24798}

\begin{abstract}
The harmonic oscillator is the paragon of physical models; conceptually and computationally simple, yet rich enough to teach us about physics on scales that span classical mechanics to quantum field theory. This multifaceted nature extends also to its inverted counterpart, in which the oscillator frequency is analytically continued to pure imaginary values. In this article we probe the inverted harmonic oscillator (IHO) with recently developed quantum chaos diagnostics such as the out-of-time-order correlator (OTOC) and the circuit complexity. In particular, we study the OTOC for the displacement operator of the IHO with and without a non-Gaussian cubic perturbation to explore genuine and quasi scrambling respectively. In addition, we compute the full quantum Lyapunov spectrum for the inverted oscillator, finding a paired structure among the Lyapunov exponents. We also use the Heisenberg group to compute the complexity for the time evolved displacement operator, which displays chaotic behaviour. Finally, we extended our analysis to N-inverted harmonic oscillators to study the behaviour of complexity at the different timescales encoded in dissipation, scrambling and asymptotic regimes.

\end{abstract}
\title{The Multi-faceted Inverted Harmonic Oscillator:\\
Chaos and Complexity}

\author{Arpan Bhattacharyya}
\email{abhattacharyya@iitgn.ac.in}
\affiliation{Indian Institute of Technology,Gandhinagar,Gujarat 382355, India}

\author{Wissam Chemissany}
\email{wissamch@caltech.edu}
\affiliation{Institute for Quantum Information and Matter, California Institute of Technology,\\
1200 E California Blvd, Pasadena, CA 91125, USA.}

\author{S. Shajidul Haque}
\email{shajidhaque@gmail.com}
\affiliation{Department of Mathematics and Applied Mathematics, University of Cape Town,
Private Bag, Rondebosch, 7701, South Africa.}

\author{Jeff Murugan}
\email{jeff.murugan@uct.ac.za}
\affiliation{Department of Mathematics and Applied Mathematics, University of Cape Town,
Private Bag, Rondebosch, 7701, South Africa.}

\author{Bin Yan}
\email{byan@lanl.gov}
\affiliation{Center for Nonlinear Studies, Los Alamos National Laboratory, Los Alamos, NM 87544, USA}
\affiliation{Theoretical Division, Los Alamos National Laboratory, Los Alamos, NM 87544, USA}

\maketitle
\tableofcontents
\section{Introduction}
One would be hard-pressed to find a physical system that we have collectively learnt more from than the harmonic oscillator. Indeed, from the simple pendulum of  classical mechanics to mode expansions in quantum field theory, there is no more versatile laboratory than the harmonic oscillator (and its many variants). This is due in no small part to two central properties of harmonic oscillator systems; they are mathematically and physically rich and simultaneously remarkably simple. It is also the universal physical response in perturbation theory.

This utility has again come into sharp relief in two seemingly disparate contexts; quantum chaos and the emerging science of quantum complexity. While neither subject is particularly new, both have seen some remarkable recent developments of late. To see why, note that conservative Hamiltonian systems come in one of two types, they are either integrable or non-integrable. The latter in turn can be classified as either completely chaotic or mixed (between chaotic, quasiperiodic or periodic), depending on whether the defining Hamiltonian is smooth or not \cite{Haake2006-ic,cvitanovic2016chaos}. By far, most non-integrable classical systems are of the latter type. The former however includes some iconic Hamiltonian systems such as the Sinai billiard model, kicked rotor and, of particular interest to us in this article, the {\it inverted harmonic oscillator} (IHO).

Classical chaotic systems are characterised by a hypersensitivity to perturbations in initial conditions under the Hamiltonian evolution. This hypersensitivity is usually diagnosed by studying individual orbits in phase space. However, as a result of the Heisenberg uncertainty principle, the volume occupied by a single quantum state in the classical phase space is $\sim \hbar^{N}$, for a system with $N$ degrees of freedom, and we no longer have the luxury of following individual orbits. This necessitates the need for new chaos diagnostics for quantum systems. One such diagnostic, discovered by Wigner in the 1950's already, is encoded in the statistical properties of energy spectra; quantum chaotic Hamiltonians have eigenvalue spacing distributions that are given by Gaussian random matrix ensembles. Unfortunately though, a direct spectral analysis of the Hamiltonian is computationally taxing for all but the simplest, or exceedingly special, systems. It therefore makes sense to develop other, complimentary diagnostics that probe different aspects of quantum chaos, say, at different times or energy scales.

One such tool, originally considered in the context of superconductivity, but rapidly gaining traction in the high energy and condensed matter communities, is the {\it out-of-time-order correlator} \cite{Kitaev2015,Larkin1969,Maldacena2016-mb}, $\mathrm{OTOC}(t)\equiv \langle B^{\dagger}(0)A^{\dagger}(t)B(0)A(t)\rangle_{\beta} $, for Heisenberg operators $A(t)$ and $B(t)$, and where $\langle\mathcal{O}\rangle_{\beta} = \mathrm{Tr}\left(e^{-\beta H}\mathcal{O}\right)/\mathrm{Tr}\,e^{-\beta H}$ denotes a thermal average at temperature $T = 1/\beta$.  One reason for this popularity is its relation to the double commutator $C_{T}(t) = -\langle [A(t),B(0)]^{2}\rangle_{\beta}$, which is the quantum analog of the classical expectation value,
\begin{eqnarray}
   \left\langle\left(\frac{\partial x(t)}{\partial x(0)}\right)^{2}\right\rangle_{\beta}\sim \sum_{n}c_{n}e^{2\lambda_{n}t}\,,
\end{eqnarray}
for a chaotic system with Lyapunov exponents $\lambda_{n}$. Indeed, it will often be more convenient to work with the double commutator instead of the four-point function OTOC$(t)$ and since, for unitary operators, the two are related through $C_{T}(t) = 2\left(1 - \mathrm{Re}\left(\mathrm{OTOC}(t)\right)\right)$, their information content, and exponential growth, is the same and are usually referred to interchangeably. In a chaotic many-body system, $C_{T}(t)$ exhibits a characteristic exponential growth from which the quantum Lyapunov exponent can be extracted. In this sense, the OTOC captures the early-time scrambling behaviour of the quantum chaotic system.

However, like any new technology, the OTOC is not without its subtlties. Among these are;
\begin{itemize}
   \item the fact that its reliability breaks down at late times, when the chaotic system starts to exhibit random matrix behaviour,
   \item a related mismatch to its classical value, where the commutator in the definition of $C_{T}(t)$ is replaced by the Poisson bracket, and
   \item no exponential growth for several single-particle quantum chaotic systems, such as the well-known stadium billiards model, or chaotic lattice systems, such as spin chains.
\end{itemize}
 All three of these points are related in some sense to Ehrenfest saturation where quantum corrections are of the same order as classical leading terms\footnote{We would like to thank the anonymous referee for their insight on this, and an earlier point. and} point to the need for a deeper understanding of the OTOC.

On the other hand, it is becoming increasingly clear that while no single diagnostic captures all the features of a quantum chaotic system, there is an emerging web of interconnected tools that offer complementary insight into quantum chaos \cite{Bhattacharyya:2019txx,Kudler-Flam2020-ew}. There is, for example, the (annealed) {\it spectral form factor} (SFF),
\begin{eqnarray}
   g(t;\beta) \equiv \frac{\langle |Z(\beta,t)|^{2}\rangle_{J}}{\langle Z(\beta,0)\rangle_{J}}\,,
\end{eqnarray}
where $Z(\beta,t)$ is the analytic continuation of the thermal partition function and the average is taken over different realizations of the system. The SFF interpolates between the essentially quantum mechanical OTOC and more standard random matrix theory (RMT) measures making it a particularly useful probe of systems transitioning between integrable and chaotic behaviour where it displays a characteristic dip-ramp-plateau shape \cite{Lau:2018kpa,Lau:2020qnl}. However, except in some special cases like bosonic quantum mechanics where it can be shown that the two-point SFF is obtained by averaging the four-point OTOC over the Heisenberg group \cite{deMelloKoch:2019rxr}, computing the SFF is a difficult task, compounded by various subtleties inherent to the spectral analysis of the chaotic Hamiltonian. 

More recently, this diagnostic toolbox has been further expanded with the introduction of a number of more information-theoretic resources with  varying degrees of utility. One of these is the {\it fidelity} \cite{Goussev2012-ep,Gorin2006-ro} of a quantum system. Let $U$ be a unitary map and $|\psi(0)\rangle$, some fiducial state in the Hilbert space. Now evolve this initial state with $U$ to $|\psi(n)\rangle = U^{n}|\psi(0)\rangle$ and again, but with a sequence of small perturbations by some non-specific field perturbation operator, to $|\widetilde{\psi}(n)\rangle = \left(e^{-iV\delta}U\right)^{n}|\psi(0)\rangle$. It was shown in \cite{PhysRevLett.89.284102} that the fidelity 
\begin{eqnarray}
   \mathcal{F}(n) = |\langle \psi(n)|\widetilde{\psi}(n)|^{2}\,,
\end{eqnarray}
for a classically chaotic quantum system, exhibits a characteristic, and efficiently computable, exponential decay under a sufficiently strong perturbation. This quantity has recently been shown \cite{Yan2020-ov,Kurchan2018-ff,Romero-Bermudez:2019vej} to be intrinsically related to the OTOC.

Closer to the focus of this article, another related tool drawn from theoretical computer science is the notion of {\it computational} (or {\it circuit}) complexity \cite{nielsen_chuang_2010}, which, in the lingo of computer science, measures the minimum number of operations required to implement a specific task in the following sense: 
fix a reference state $|\Psi_{\mathrm{R}}\rangle$ and target state $|\Psi_{\mathrm{T}}\rangle$ and construct a unitary $U$ from a set of elementary gates by sequential operation on the reference state such that $|\Psi_{\mathrm{T}}\rangle = U |\Psi_{\mathrm{R}}\rangle$. Then the complexity of $|\Psi_{\mathrm{T}}\rangle$ is defined to be the minimal number of gates required to implement the unitary transformation from reference to target states. Determining the computational complexity is then essentially an optimisation problem, one that was more or less solved by Nielsen in \cite{NL1}. Nielsen's geometrical approach proceeds by defining a {\it cost functional}
\begin{eqnarray}
   \mathcal{D}[U(t)] \equiv \int_{0}^{1} dt\, F\left(U(t),\dot{U}(t)\right)\,
\end{eqnarray}
on the space of unitaries which is then optimised subject to the boundary conditions $U(0)=1\!\!1$ and $U(t)=U$. For a long time, the idea of circuit complexity was viewed as a curiosity of computer science, living on the periphery of theoretical physics. This situation changed dramatically with the introduction, by Susskind and collaborators \cite{Susskind:2014rva,Stanford:2014jda,Brown:2015bva}, of complexity as a probe of black hole physics. Further, the idea of complexity has  extended to quantum field theory in recent time  \cite{MyersCC,Chapman:2017rqy,Caputa:2017yrh,Bhattacharyya:2018bbv,Hackl:2018ptj,Khan:2018rzm,Camargo:2018eof,Ali:2018aon,Bhattacharyya:2018wym,Caputa:2018kdj,Bhattacharyya:2019kvj,Caputa:2020mgb,Flory:2020eot,Erdmenger:2020sup, Bhattacharyya:2020rpy,Bhattacharyya:2020kgu,DiGiulio:2020hlz,Caceres:2019pgf,Susskind:2020gnl,Chen:2020nlj,Czech:2017ryf,Camargo:2019isp,Chapman:2018hou,Chapman:2019clq,Doroudiani:2019llj,Geng:2019yxo,Guo:2020dsi,Guo:2020dsi,Guo:2020dsi}. Following this line of reasoning, two disjoint subsets of the current authors conjectured that not only does the computational complexity furnish an equivalent chaos diagnostic to the OTOC \cite{me1,Balasubramanian:2019wgd,Yang:2019iav,Yang:2019udi,Yang:2019udi}, but in addition, a response matrix may be defined to characterize the fine structure of the complexity in response to initial perturbations, giving rise to the full Lyapunov spectrum in the classical limit \cite{Yan:2020twr}. Table I summarises the findings of these studies and compares the time development of the complexity to that of the OTOC. It is worth emphasizing that the universal behaviors summarized in Table I are for generic operators and complex chaotic systems; one can always cook up special scenarios which are not described by these generic forms. For instance, as shown in the following sections, the OTOC of the IHO for the displacement operators does not decay at all, whereas the OTOC for canonical variables, i.e., $x$ and $p$, the OTOC exhibits intermediate exponential decay. The early scrambling regime of the OTOC will not be discussed in the work, it has been demonstrated for coupled IHOs in Ref.~\cite{Yan2020-ov}. We also stressed that the presence of the early scrambling is limited, i.e., models exhibiting this regime usually show a large hierarchy between scrambling and local dissipation time scales \cite{Maldacena2016-mb}. Many chaotic systems, e.g., spin chains \cite{Craps2020-oq}, only manifest a pure exponential intermediate decay. On the other hand, our study of the complexity reveals both the early and intermediate regimes, as well as the local dissipation before the scrambling time scale (Fig.~\ref{fig_compgrowth}).

\begin{table}[h!]
    \centering
\begin{tabular}{ |c|c|c| } 
 \hline
  & Early scrambling  & Intermediate regime  \\ 
  \hline 
 OTOC & $1 - \epsilon e^{\lambda t}$ ~$\sim \exp(-\epsilon e^{\lambda t})$ &  $e^{- \Gamma t}$ \\ 
 \hline
 Complexity & $\epsilon e^{\lambda t}$ & $\Gamma t$ \\ 
 \hline
\end{tabular}
\caption{Universal correspondence between OTOC and complexity, complexity $\sim -\log$(OTOC). This relation holds at both the early scrambling and intermediate decay regime.}
\end{table}

In both cases, the system chosen to exhibit this relationship between the complexity and quantum chaos was arguably among simplest conceivable; the inverted harmonic oscillator.
 The inverted harmonic oscillator does not resemble typical large N chaotic systems in that the decay of the OTOC, or the growth in the complexity, here captures an instability of the system, rather than chaos.Nevertheless, it remains a useful toy model with which to study the various chaos diagnostics.
The present article builds on these ideas by returning to the inverted oscillator, developing the treatment of the OTOC as well as the computational complexity of particular states in the model and then connecting them. In addition to its pedagogical value the oscillator again provides a rich and intuitively clear example within which to understand further the OTOC and computational complexity. It is therefore fitting that we begin with a brief overview of the inverted harmonic oscillator.

\section{The IHO Model}
We start with the harmonic oscillator Hamiltonian
\begin{equation}
    H=\frac{p^{2}}{2m}+\frac{m\omega^{2}}{2}x^{2},
\end{equation}
where $p \equiv -i\hbar\frac{d}{dx}$ is the momentum operator. We will work in natural units in which $\hbar = 1$ and, without any loss of generality, assume that the mass of the oscillator $m=1$. By choosing the value of the frequency $\omega$, three different cases can be obtained:
\begin{equation*}
    \omega=
        \begin{cases}
    &\Omega\quad\textrm{harmonic oscillator,}\\
    &0\quad\textrm{free particle,}\\
    &i\Omega\quad\textrm{inverted harmonic oscillator.}
    \end{cases}
\end{equation*}
Here $\Omega$ is a positive real number. In this work, we will be mainly concerned with the Hamiltonian of the inverted harmonic oscillator. \\

An important point to note is that the regular and inverted harmonic oscillators are genuinely different. As a result, one cannot take for granted that formulae known for the regular oscillator and extrapolate them to the inverted oscillators by simply replacing $\Omega$ with $i\Omega$. For instance, the regular harmonic oscillator has a discrete energy spectrum $(n+1/2)\Omega$, while the spectrum for an inverted oscillator is a continuum. However, in some other cases, such as the Heisenberg evolution for the position or momentum operator, the derivation follows in much the same way for both the regular and inverted oscillators. In such cases, we will explicitly point this out and use the variable $\omega$ to cover both classes of oscillator. It will be useful in what follows, to define the annihilation and creation operators \cite{bermudez2013factorization}, 
\begin{equation}
    a^{\pm}_{\omega}=\frac{1}{\sqrt{2}}\left(\mp ip+\omega x\right).
\end{equation}

%
%
%
\section{Out-of-Time Order Correlator}
We will consider the OTOC for the displacement operators in the IHO, which are defined in terms of the creation and annihilation operators as
\begin{equation}\label{DO}
    D\left(\alpha\right)=\exp\left(\alpha a^{\dagger}-\alpha^{*}a\right).
\end{equation}
This is a well known operator whose phase space OTOC has been the subject of recent study for continuous variable (CV) systems \cite{zhuang2019scrambling}. There the authors argued that, for a Gaussian-CV system, the OTOC does not display any genuine scrambling\footnote{
 An initial operator will be said to be {\it genuinely} scrambling (non-Gaussian) when it is localized in phase space and spreads out under time evolution of the system. A local ensemble of operators is said to be quasi scrambling (Gaussian) when it distorts but the overall volume of the phase space remains fixed.}. In this section, we would like to explicitly check this claim for the IHO. Since the Hamiltonian of the IHO belongs to the category of Gaussian-CV systems, we expect the OTOC for the displacement operators to display such quasi-scrambling.  More explicitly, the OTOC only changes by an overall phase, while the magnitude remains constant. We follow this by adding a cubic-gate perturbation to the oscillator potential and explore the OTOC analytically. In contrast to its Gaussian counterpart, this model does indeed display generic scrambling behavior.\\

In the second part of this section, we will investigate another important feature of chaos; the Lyapunov spectrum.  Our goal will be to use the IHO to check whether the quantum Lyapunov exponents exhibit a similar pairing structure to the classical case.





\subsection{OTOC for the Displacement Operator}
To warmup, we will evaluate the OTOC for the displacement operator for the regular harmonic oscillator with real frequency $\omega =\Omega$. Our derivation is esentially independent of the choice of the frequency $\omega$. Therefore, by choosing appropriate values of the frequency, we derive expressions for both the regular and inverted harmonic oscillator. The displacement operator in Eq.(\ref{DO}) for a single mode harmonic oscillator can be written as
\begin{equation}\label{eq:Doperator}
    D(\alpha)=\exp\left[i\sqrt{2}\left(\text{Im}(\alpha) \omega x- \text{Re}(\alpha)p\right)\right].
\end{equation}
To evaluate the OTOC, we need to find the time evolution of the displacement operator (\ref{eq:Doperator}), i.e.,
\begin{equation}
    D(\alpha,t)=e^{iHt}D(\alpha,t=0)\ e^{-iHt},
\end{equation}
which can be evaluated by implementing the Hadamard lemma. Some straightforward algebra puts this into the form,
\begin{equation} \label{dis}
\begin{split}
    D(\alpha,t)    &=\exp\big[i\sqrt{2}\left(\text{Im}(\alpha)\cos(\omega t)+ \text{Re}(\alpha)\sin(\omega t)\omega\right) x\\
    & \qquad +i\sqrt{2}\left(\text{Im}(\alpha)\sin(\omega t)/\omega-\text{Re}(\alpha)\cos(\omega t)\right)p\big].
\end{split}
\end{equation}
The corresponding OTOC function, $C_2(\alpha,\beta;t)_\rho$ is defined as
\begin{equation}
\begin{split}\label{OTOC}
    C_2(\alpha,\beta;t)_\rho&\equiv \langle D^{\dagger}(\alpha,t)D^{\dagger}(\beta)D(\alpha,t)D(\beta)\rangle\\
    &= \text{Tr}[\rho D^{\dagger}(\alpha,t)D^{\dagger}(\beta)D(\alpha,t)D(\beta)],\\ 
\end{split}
\end{equation}
where $\rho$ is a given state of the harmonic oscillator.
By using (\ref{dis}), the OTOC (\ref{OTOC}) simplifies to the following form
\begin{equation}\label{OTOC-HO}
    C_2(\alpha,\beta;t)_\rho=\exp(i\theta),
\end{equation}
where 
\begin{equation}
    \theta=2\omega \text{Re}(\alpha\beta^{\ast})\sin(\omega t)+2 \text{Im}(\alpha\beta^{\ast})\cos(\omega t).
\end{equation}
We can immediately see that $\theta$ is real-valued for both $\omega=\Omega$ and $\omega=i\Omega$ and we conclude that the magnitude of the OTOC  (\ref{OTOC-HO}) does not decay in time for either the regular or inverted harmonic oscillators. This implies that the harmonic oscillator potential is quasi-scrambling, in agreement with the general conclusion for the Gaussian dynamics found in \cite{zhuang2019scrambling}. 
%
%
%
We can extend the IHO Hamiltonian to a simple non-Gaussian one by adding a so-called cubic-gate as follows:
%
\begin{equation*}
   H=\frac{p^2}{2m}+ \frac{m \omega^2}{2} x^2 +\gamma\frac{x^{3}}{3!}\quad 
\end{equation*}
To display more clearly the role of the different contributions in the Hamiltonian to the OTOC we rewrite the Hamiltonain in the following form
\begin{equation*}
   H= k p^2 + l x^2 +J x^3,
\end{equation*}
where $k = \frac{1}{2m},l= \frac{m \omega^2}{2}$ and $\frac{\gamma}{3!} = J$. 
As in the previous Gaussian case, we first find the time evolution of the displacement operator (\ref{eq:Doperator}) for this cubic model. It has the following form
    %
\begin{equation}
\begin{split}
    D(\alpha,t)
    &=\exp\Big[i\Big(A_0+A_1 p+A_2 p^2+A_3 x+A_4 x^2 \\
    &\qquad +\mathcal{O}(x^3,p^3,px,t^3)\Big)\Big]
\end{split}
\end{equation}
where $A_i$'s are functions of $k,\, l\, ,J$ and $t.$
From this expression, the OTOC can be computed exactly as
\begin{equation}
    C_2(\alpha,\beta;t)_\rho=\exp \left (i\theta (k, l, J)\right)\ \chi(12i J t \text{Re}(\alpha) \text{Re}(\beta),\rho),
\end{equation}
where $\chi(12 i J t \text{Re}(\alpha) \text{Re}(\beta),\rho)$ is the characteristic function \cite{dangniam2015quantum}, which typically decays in time. To illustrate this, consider a thermal state ${\rho}_{n_{th}}$ for which the characteristic function has the form
\begin{equation}
\chi(\xi,\rho_{n_{th}})=\exp\left[-(n_{th}+\frac{1}{2})|\xi|^{2}\right].
\end{equation}
Using this, the OTOC reads
\begin{multline}\label{Otcubic}
    C_2(\alpha,\beta;t)_{{\rho}_{n_{th}}}=\exp(i\theta)\ \chi(2i\gamma t \text{Re}(\alpha) \text{Re}(\beta),\rho_{n_{th}})\\
    =\exp(i\theta)\ \exp \left [-2(2n_{th}+1)\left(\text{Re}(\xi)^2 + \text{Im}(\xi)^{2} \right ) \right],
\end{multline}
where 
\begin{equation} \label{cubic}
    \begin{split}
   \text{Re}(\xi)  &= - 6 k J \text{Re} (\alpha) \text{Re} (\beta) t^2, \\
   \text{Im}(\xi)  &= - 6 J \text{Re} (\alpha) \text{Re} (\beta) t - 6 J k \text{Im} (\alpha \beta^*)  t^2.
   \end{split}
\end{equation}
We can immediately see that there will be an exponential decay when the cubic term is added to the Hamiltonian. This essential role of the cubic term is clear from Eq.~(\ref{Otcubic}) and Eq.~(\ref{cubic}).  More precisely, the overall minus sign of the exponent in the amplitude of eq.~(\ref{Otcubic}) signals a Gaussian decay in time. When $J=0$ (and $k \neq 0)$, the entire exponent vanishes, and the amplitude becomes unity. This is true for any value of k. On the other hand, when $J\neq 0$ and $k=0$ we still get exponential decay. This is the case discussed in \cite{zhuang2019scrambling}. Finally when both $J \neq 0, k \neq 0$, we get contributions from both of them, namely from the $x^3$ and $p^2$ terms. However, the contribution coming from the $p^2$ will never cancel the contribution coming from the cubic term. We have also checked that even if we add the higher order $A_i$'s, our conclusion, namely the decaying of the OTOC,  remains intact.  It is worth emphasising that our analysis, namely the structure and the behaviour of the derived equations, are independent of the fact whether the oscillator inverted or regular. 

\subsection{Quantum Lyapunov Spectrum}

A defining property of a classical chaotic system is its hyper-sensitivity to initial conditions in the phase space. This manifests in the exponential divergence of the distance between two initially nearby trajectories. The rate of this divergence is encoded in the so-called Lyapunov exponent. Technically, this is only the largest of a sequence of such exponents that constitute the Lyapunov spectrum. For a $2n$-dimensional phase space, the Lyapunov spectrum consists of $2n$ characteristic numbers captured by the eigenvalues of the Jacobian matrix
\begin{equation}
    M_{ij}(t) \equiv \frac{\partial z_i(t)}{\partial z_j},
\end{equation}
where the $z_i$ are phase space coordinates. The eigenvalues $s_i(t)$ of the Jacobian matrix evolve exponentially in time and the Lyapunov spectrum can is extracted in the asymptotic limit,
\begin{equation}
    \lambda_i \equiv \lim_{t\rightarrow\infty} \frac{1}{t} \ln s_i(t).
\end{equation}
If the initial perturbation in the phase space is applied to the ``eigen-direction'' with respect to one eigenvalue in the Lyapunov spectrum, the trajectories diverge with a corresponding exponential rate. For generic perturbations which involve all exponents in the Lyapunov spectrum, in the asymptotic limit, the exponential with the largest exponent will eventually dominate. In this case only the maximum Lyapunov exponent is visible.\\ 

It is worth emphasizing that the Lyapunov spectrum is typically computed from the eigenvalues of the matrix
\begin{equation}\label{eq:lmatrix}
    L(t)\equiv M(t)^\dag M(t).
\end{equation}
Due to the intrinsic symplectic structure of the classical phase space, the $M$-matrix is symplectic and, hence, the Lyapunov exponents always come in pairs with opposite signs.\\ 

Now let's think about quantum systems. In form of the commutator square, the OTOC typically grows exponentially, with a rate analogous to the classical Lyapunov exponent, i.e., for generic operators,
\begin{equation}
    \langle [W(t),V]^2  \rangle \sim \epsilon e^{\lambda t},
\end{equation}
up-to the time scale known as the scrambling (or Ehrenfest) time \cite{Maldacena2016-mb}. For systems with a classical counterpart, e.g., quantum kicked rotor, the growth rate of the OTOC indeed matches the \emph{maximum} classical Lyapunov exponent. A natural question to ask is if it possible to fine-tune the operators in the OTOC and extract a full spectrum of Lyapunov exponents, instead of only the leading one? Some recent attempts \cite{Gharibyan2019-sp,Grozdanov:2018atb} to tackle this problem generalized the Jacobian matrix to quantum systems using the OTOCs:
\begin{equation}\label{eq:mmatrix}
    M_{ij}(t)\equiv i[z_i(t),z_j].
\end{equation}
Here $z_i$ ranges over canonical variables, and $z_i(t)$ is the Heisenberg evolution. In contrast to the classical case, the quantum instability matrix (\ref{eq:mmatrix}) lacks a symplectic structure. Known examples such as spin chains and the finite size $SYK$-model show that the quantum Lyaponov exponents do {\it not} come in pairs \cite{Gharibyan2019-sp}. However, these models lack any well-behaved exponential growth in the first place. \\

For the IHO, the Heisenberg evolution of the canonical pair of variables $\{x,p\}$ can be computed exactly,
\begin{equation}
    \begin{aligned}
        x(t)&=x(0)\cosh{\Omega t}+\frac{1}{m\Omega}p(0)\sinh{\Omega t}\\
        p(t)&=p(0)\cosh{\Omega t}+m\Omega x(0)\sinh{\Omega t}.
    \end{aligned}
\end{equation}
This in turn allows us to compute the quantum Jacobian matrix,
\begin{equation}
     M(t)=
     \begin{pmatrix}
    \sinh{\Omega t}/(m\Omega) & -\cosh{\Omega t}\\
    \cosh{\Omega t} & -m\Omega\sinh{\Omega t}
        \end{pmatrix},
\end{equation}
the elements of which are OTOCs that all grow exponentially with the largest Lyapunov exponent $\Omega$. Once we diagonalize the above matrix, the hyperbolic functions arrange in such a way that a pair of exponentials emerge with exponents $\pm\Omega$. This coincides with the Lyapunov exponent of the classical inverted oscillator.  
%
\section{Complexity for Inverted Harmonic Oscillator}
Recently by using the inverted harmonic oscillator, {\it complexity} has been proposed as a new diagnostic of quantum chaos \cite{me3,Bhattacharyya:2019txx}. Depending on the setup and details of the quantum circuit, the scope and sensitivity of computational complexity as a diagnostic can vary. Since this is a fairly new diagnostic, its full capacity to capture chaotic behaviour is not yet fully understood. Therefore, to gain further insight into quantum chaotic systems, we will extend our investigation in two different directions. 
First, we compute the complexity for the displacement operator by using the operator method of Nielsen \cite{NL1,NL2,NL3,MyersCC}, which explores how one can construct a given operator from the identity. Then we will develop a construction based on the Heisenberg group, which provides a natural basis choice for the displacement operator. Notice that this displacement operator description is analogous to the doubly evolved quantum circuit constructed in Ref.~\cite{me3}. We would like to explore if this operator formalism is consistent with the existing results and whether it can provide us with any additional information about chaos.\\

In the second part of this section, we will use the wave function, or correlation matrix method, for a system of N-oscillators to study the behaviour of complexity at different timescales, namely, dissipation, scrambling and asymptotic regimes. This particular setup introduces two new parameters into the problem: the number of oscillators and the lattice spacing, and we will also determine how complexity and the scrambling time depends on them.
\subsection{Complexity for the Displacement Operator }
Now let's compute the circuit complexity corresponding to the time evolved displacement operator. We evolve the displacement operator mentioned in (\ref{eq:Doperator}) by the inverted harmonic oscillator Hamiltonian. We get the following,
\begin{equation}
\label{op1}
D(\alpha,t)= \text{exp} \big [ A(t)\, i\, x + B(t)\, i\, p\big],
\end{equation}
where
\begin{equation}
\begin{split}
A(t)&=\sqrt{2}\left(\text{Im}(\alpha)\cosh(\Omega t)-\text{Re}(\alpha)\sinh(\Omega t)\Omega\right),\\
B(t)&=\sqrt{2}\left(\text{Im}(\alpha)\sinh(\Omega t)/\Omega-\text{Re}(\alpha)\cosh(\Omega t)\right).
\end{split}
\end{equation}

Evidently it is an element of the Heisenberg group. Consequently, we can parametrize the unitary as,
\begin{equation} \label{unitary}
    U(\tau)= {\overleftarrow{\mathcal{P}}}\exp(i\int_{0}^{\tau }\, d\tau\, H(\tau)),
\end{equation}
where,
\begin{equation}
    H(\tau)=\sum_{a} Y^a (\tau) O_a.
\end{equation}
$O_a=\{i\,x, i\,p, -i\,\hbar\,I \}$ generators of Heisenberg group whose algebra is defined by,
\begin{equation} \label{alg}
[i\,x,i\,p]=-i\hbar\, I,[i\,x, -i \,\hbar\, I]=0, [i\,p,-i\,\hbar\, I]=0.   \end{equation}
The associated complexity function is defined as, 
\begin{equation}\label{func}
    \mathcal{C}(U)=\int_0^1 d\tau\sqrt{G_{ab}Y^{a}(\tau) Y^{b}(\tau)}.
\end{equation}
We choose $G_{ab}=\delta_{ab}.$
To proceed further we choose to work with following representation of Heisenberg generators. For ease of computation, we start with the 3-dimensional representation of the Heisenberg group generators \cite{Hall:2000xt}.
\begin{equation} \label{3by3}
    M_1 = \begin{pmatrix}
0 & 1 & 0\\
0 & 0 & 0\\
0 & 0 & 0
\end{pmatrix},
M_2 = \begin{pmatrix}
0 & 0 & 0\\
0 & 0 & 1\\
0 & 0 & 0
\end{pmatrix},
M_3 = \begin{pmatrix}
0 & 0 & 1\\
0 & 0 & 0\\
0 & 0 & 0
\end{pmatrix}
\end{equation}
It can be easily checked that these $M_a$'s satisfy the same algebra as that of (\ref{alg}). 
From (\ref{unitary}), and using the expressions of $M_a$'s we can easily show that,
\begin{equation}
  Y^{a}=\text{Tr}(\partial_{\tau} U(\tau). U^{-1}(\tau). M_a^{T}).
    \end{equation}
This in turn helps us to define a metric on this space of unitaries,
\begin{equation} \label{met}
\begin{aligned}
        ds^2= \delta_{ab}(\text{Tr}(\partial_{\tau} U(\tau). & U^{-1}(\tau). M_a^{T}))\times\\
        &\times (\text{Tr}(\partial_{\tau} U(\tau). U^{-1}(\tau). M_b^{T}))
\end{aligned}
\end{equation}
The last step is to minimize the complexity functional (\ref{func}). The minimum value that it takes will then give the required complexity. It can be shown, following \cite{MyersCC,Guo:2018kzl,Bhattacharyya:2018bbv}, that (\ref{func}) can be minimized by evaluating it on the geodesics of (\ref{met}) with the boundary conditions,
\begin{equation} \label{bndy}
    \tau=0, U(\tau=0)=I, \tau=1, U(\tau=1)=D(\alpha,t),
\end{equation}
where in the representation (\ref{3by3}) $D(\alpha,t)$ becomes,
\begin{equation} \label{final}
D(\alpha,t)=\begin{pmatrix}
1& A(t) & \frac{1}{2} A(t) B(t)\\
0 & 1 & B(t)\\
0 & 0 & 1
\end{pmatrix}.
\end{equation}
For our case $U(\tau)$ is an element of Heisenberg group so we can parametrize $U(\tau)$ as,
\begin{equation}
U(\tau)=\begin{pmatrix}
1 &x_1(\tau) & x_3(\tau)\\
0 & 1 & x_2(\tau)\\
0 & 0 & 1
\end{pmatrix},
\end{equation}
and, given this parametrization, from (\ref{met}) we find that,
\begin{equation}
ds^2=(1+x_2^2) dx_1^2+ dx_2^2+ dx_3^2-2 x_2\, dx_1 dx_3,
\end{equation}
and the complexity functional,
\begin{equation}\label{func1}
    \mathcal{C}(U)=\int_0^1 d\tau\sqrt{g_{ij}\dot x^{i}(\tau) \dot x^{j}(\tau)},
    \end{equation}
where, $x^{i}=\{ x_1, x_2, x_3\}.$  We have to minimize (\ref{func1}) using the boundary conditions (\ref{bndy}). For this we need  to  solve  for the geodesics of this background, which amounts to solving second order differential equations. Alternatively, we can find the killing vectors of this space and the corresponding conserved charges to formulate the system as a first order one. We first list the Killing vectors below,
\begin{align}
\begin{split}
&k_1=\frac{\partial}{\partial x_1},\\&
k_2=\frac{\partial}{\partial x_2}+x_1\frac{\partial}{\partial x_3},\\&
k_3=\frac{\partial}{\partial x_3}.
\end{split}
\end{align}
The corresponding conserved charges ($c_I= (k_I)^i g_{ij} \dot x^{j}$) are,
\begin{align}
\begin{split} \label{eq}
&c_1=(1+x_2(\tau)^2)\dot x_1(\tau)-x_2(\tau) \dot x_3(\tau),\\&
c_2=\dot x_2(\tau)+x_1(\tau)\dot x_3(\tau)-x_1(\tau)x_2(\tau)\dot x_1(\tau),\\&
c_3=\dot x_3(\tau)-x_2(\tau)\dot x_1(\tau).
\end{split}
\end{align}
To solve these first order differential equations we first set $c_3=0$ in (\ref{eq}) to get,
 \begin{align}
 \begin{split}
 &\dot x_{1}(\tau)=c_1,\\&
 \dot x_2(\tau)=c_2,\\&
 \dot x_3(\tau)=x_2(\tau) \dot x_1(\tau).
 \end{split}
 \end{align}
  The solutions for these equations are,
  \begin{align}
  \begin{split}
  &x_{1}(\tau )= \chi_1+c_{1} \tau,\\&
  x_{2}(\tau )= \chi_2+c_{2} \tau ,\\&
  x_{3}(\tau )=\chi_3+ c_1 \chi_{2} \tau +\frac{1}{2} c_{1} c_{2} \tau ^2
  \end{split}
  \end{align}
   From $\tau=0$ boundary condition, $\chi_1=\chi_2=\chi_3=0.$
     Then we are left with,
    \begin{align}
  \begin{split}
      &x_{1}(\tau )= c_{1} \tau,\\&
  x_{2}(\tau )= c_{2} \tau ,\\&
  x_{3}(\tau )= \frac{1}{2} c_{1} c_{2} \tau ^2
  \end{split}
  \end{align}
  Then from the final boundary condition at $\tau=1$ we have,
  \begin{equation}
  c_1=A(t), c_2=B(t).
  \end{equation}
  Then finally we have,
  \begin{equation}
  x_1(\tau)=A(t)\tau, x_2= B(t)\tau, x_3(\tau)=\frac{1}{2} A(t) B(t) \tau^2.
  \end{equation}
 We evaluate the complexity with this  solution,
\begin{equation}
\mathcal{C}(U)=\sqrt{A(t)^2+B(t)^2}.
\end{equation}
This is a remarkably simple expression. We suspect that this is a consequence of the simple structure of the Heisenberg group. One can immediately see the behaviour of complexity at large times where it grows as a simple exponential
\begin{equation}
    \mathcal{C}(U) \approx \frac{\text{Im}(\alpha)-\text{Re}(\alpha)\Omega}{\sqrt{2}\Omega}(\sqrt{1+\Omega^2})e^{\Omega t}.
\end{equation}
\begin{figure}[t]
\includegraphics[width=0.45 \textwidth]{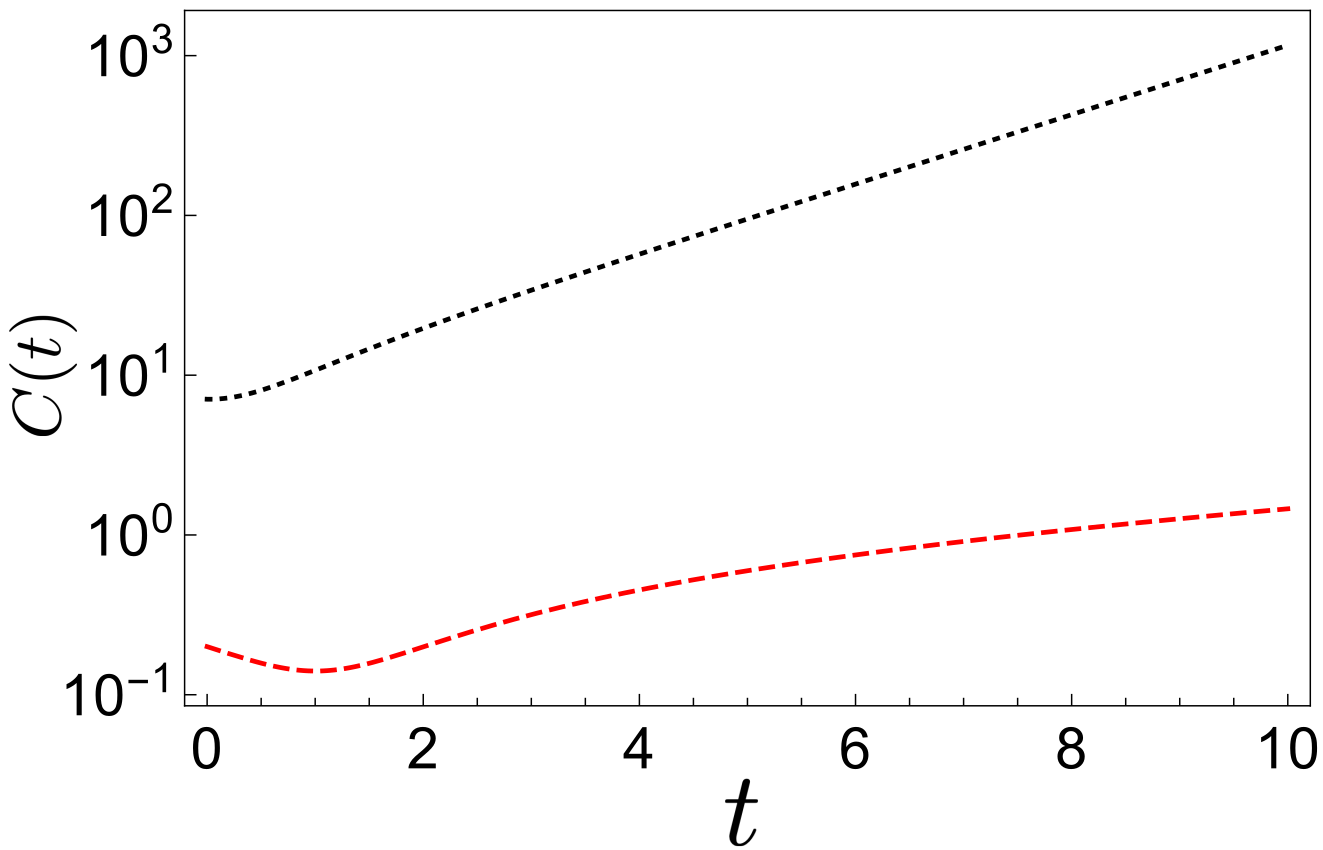}
\caption{Time evolution of complexity of the displacement operator (computed from the operator method), for different choice of parameters. The red and the blue dotted curves correspond to $\{ \text{Im} (\alpha) = 0.1, \text{Re} (\alpha) = 0.1, \Omega = 0.1 \}$ and $\{ \text{Im} (\alpha) = 5, \text{Re} (\alpha) = 0.1, \Omega= 0.5  \}$ respectively}
\label{figoperator}
\end{figure}
Fig.~\ref{figoperator} displays the time evolution of complexity for two different sets of parameters in the logarithmic scale. Note that the overall behaviour is chaotic as expected for IHO and it matches with Ref.~\cite{me3}. The late time behavior for both cases are exponential as expected. The early times behaviour on the other hand is a bit subtle. 
Looking closer, we notice that for a particular set of values of the parameters there is a minimum in the evolution of complexity during the scrambling-time regime. This strange feature is absent for the complexity computed from the correlation matrix method used in Ref.~\cite{me3}. The physical significance of this minimum is unclear to us at this point. To understand its implications and whether it is a generic feature for this method will require further investigations of other models. Naively though, it hints that the operator method is perhaps more sensitive than the wave function method for computing complexity. We would like to investigate these issues elsewhere. Also, we note that conclusion drawn here is not sensitive to the choice of the cost functional (\ref{func1}). One could have certainly chosen another cost functional. For a detailed discussion of the various choices of interested readers are referred to \cite{Guo:2018kzl}.

\subsection{Complexity for N-Oscillators and Scrambling}
To further investigate the scrambling behaviour for the inverted harmonic oscillator, in this sub-section we will take a different approach. First of all we will compute the state complexity instead of operator complexity. Secondly, we will consider a large number of inverted harmonic oscillators. To establish our point we will use the model used by Ref.~\cite{me1}, where the authors extended the inverted harmonic oscillator model and considered the field theory limit. Below we start with a review of the model studied in Ref.~\cite {me1}. 

First we will consider two free scalar field theories ((1+1)-dimensional $c=1$ conformal field theories) deformed by a marginal coupling as in Ref.~\cite {me1}. The Hamiltonian for such model is given by
\begin{equation}\label{eq:field}
\begin{aligned}
H &= H_0+H_I= \frac{1}{2}\int dx \Big[\Pi_{1}^2+(\partial_{x}\phi_1)^2+\Pi_{2}^2+(\partial_{x}\phi_2)^2 \\
&+m^2 (\phi_1^2+\phi_2^2)\Big]+ \lambda \int dx (\partial_{x}\phi_1)(\partial_{x}\phi_2).
\end{aligned}
\end{equation}
We will discretize this theory by putting it on a lattice. Using the following re-definitions 
\begin{equation}
\begin{aligned} 
&x(\vec n)=\delta \phi(\vec n), \ p(\vec n)=\Pi(\vec n)/\delta, \ \omega=m, \\&
\Omega =\frac{1}{\delta^2},\,  \hat \lambda=\lambda\, \delta^{-4} \ \text{and}\ \hat m=\frac{m}{\delta},
\end{aligned}
\end{equation}
we get the following Hamiltonian
\begin{equation}
\begin{aligned}
H &=\frac{\delta }{2}\sum_{n} \Big[p_{1,n}^2+p_{2,n}^2 +\Big(\Omega^2\,(x_{1,n+1}-x_{1,n})^2 \\ &+\Omega^2\,(x_{2,n+1}-x_{2,n})^2+\big(\hat m^2( x_{1,n}^2+ x_{2,n}^2) \\ 
&+ \hat \lambda\, (x_{1,n+1}-x_{1,n})(x_{2,n+1}-x_{2,n})\Big)\Big].
\end{aligned}
\end{equation}
Next we perform a series of transformations,
\begin{align}
\begin{split}
x_{1,a} &= \frac{1}{\sqrt{N}}\sum_{k=0}^{N-1} \exp\Big(\frac{2\pi\,i\,k}{N}\, a\Big)\tilde x_{1,k},\\
p_{1,a} &= \frac{1}{\sqrt{N}}\sum_{k=0}^{N-1} \exp\Big(-\frac{2\,\pi\,i\,k}{N}a\Big)\tilde p_{1, k},\\
x_{2,a} &= \frac{1}{\sqrt{N}}\sum_{k=0}^{N-1} \exp\Big(\frac{2\pi\,i\,k}{N}\, a\Big)\tilde x_{2,k},\\
p_{2,a} &= \frac{1}{\sqrt{N}}\sum_{k=0}^{N-1} \exp\Big(-\frac{2\,\pi\,i\,k}{N}a\Big)\tilde p_{2, k},\\
\tilde p_{1,k} &= \frac{p_{s,k}+p_{a,k}}{\sqrt{2}},\ \tilde p_{2,k}=\frac{p_{s,k}-p_{a,k}}{\sqrt{2}},\\
\tilde x_{1,k} &= \frac{x_{s,k}+x_{a,k}}{\sqrt{2}},\ \tilde x_{2,k}=\frac{x_{s,k}-p_{a,k}}{\sqrt{2}},
\end{split}
\end{align}
that lead to the Hamiltonian
\begin{align}
\label{Ham1}
H=\frac{\delta}{2}\sum_{k=0}^{N-1}&\Big[p_{s,k}^2+\bar \Omega_k^2  x_{s,k}^2+p_{a,k}^2+\Omega_k^2 x_{a,k}^2\Big], 
\end{align}
where 
\begin{align}
\begin{split}
\bar \Omega_k^2= \left(\hat m^2+4\,(\Omega^2+\hat \lambda)\,\sin^2\Big(\frac{\pi\,k}{N}\Big)\right), \\
\Omega_k^2= \left(\hat m^2+4\,(\Omega^2-\hat \lambda)\,\sin^2\Big(\frac{\pi\,k}{N}\Big)\right). 
\end{split}
\end{align}
\begin{figure}[t]
\includegraphics[width=0.45 \textwidth]{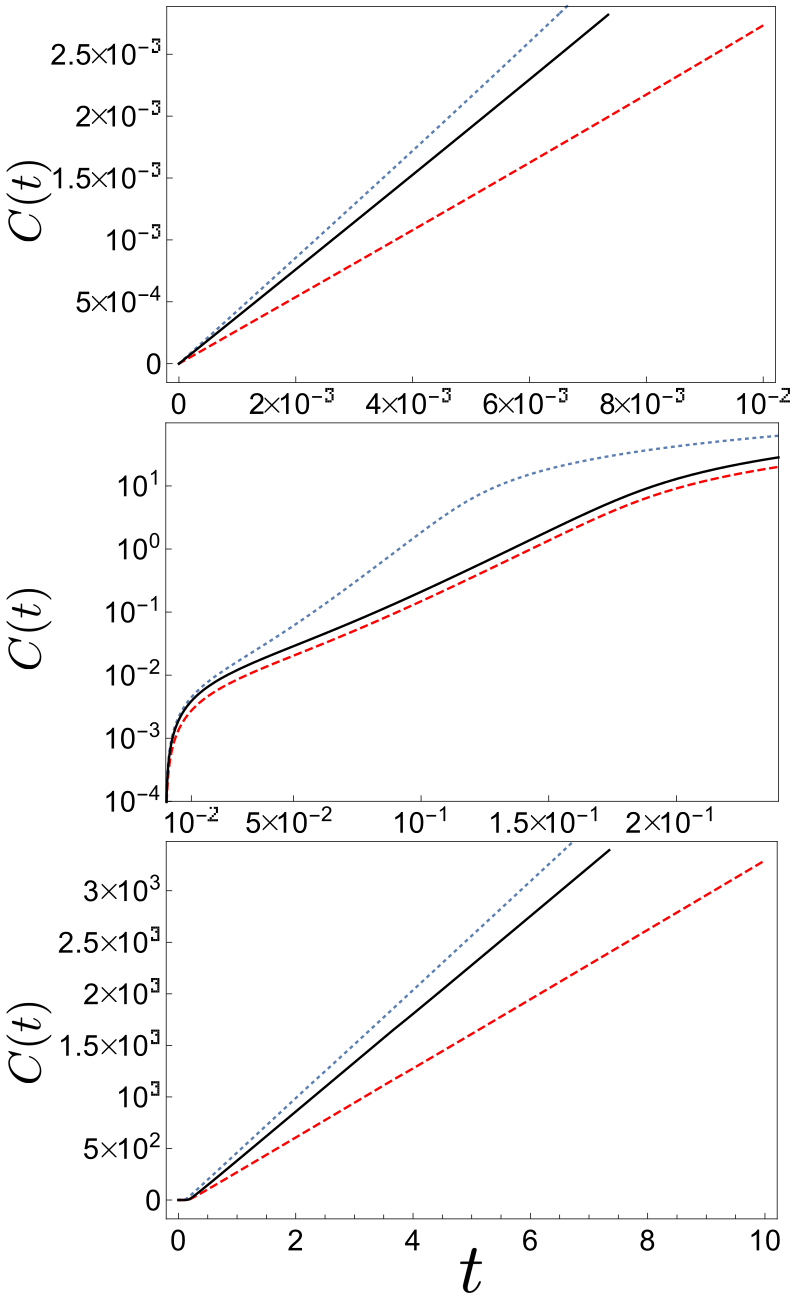}
\caption{Universal growth of the complexity in Eq.~(\ref{eq:complexityN}) at different time scales. Top, middle, and bottom figures show, respectively, the power-law dissipation, the exponential scrambling in semi-log scale, and the intermediate linear growth. Blue dotted, black, and red dashed curves correspond to \{$\delta= 0.4$, $N=100$\}, \{$\delta= 0.5$, $N=200$\}, and \{$\delta= 0.5$, $N=100$\}, respectively. Other parameters are fixed as $m=1$, $\lambda=10$, $ \delta \lambda=0.01$.}
\label{fig_compgrowth}
\end{figure}
Note that the underlying model of interest is still the inverted harmonic oscillators. It becomes immediately clear by appropriately tuning the value of $\hat \lambda$--the frequencies $\Omega_k$ can be made arbitrarily negative resulting in coupled inverted oscillators. The other frequency, $\bar \Omega_k$, however, will be always positive. Therefore, effectively this model (\ref{Ham1}) can be seen as the sum of a regular and inverted oscillator for each value of $k$. Since we are interested in the inverted oscillators we will ignore the regular oscillator part and we will simply use the inverted oscillator part of the Hamiltonian
\begin{align}
\begin{split} \label{Ham2}
\tilde H (m,\Omega,\hat \lambda)=\\\frac{\delta}{2}\sum_{k=0}^{N-1}&\left[p_{k}^2+\left(\hat m^2+4\,(\Omega^2-\hat \lambda)\, \sin^2\left(\frac{\pi\,k}{N}\right) \right)x_{k}^2\right].
\end{split}
\end{align}
Even with this Hamiltonian, by tuning $\hat\lambda$ we can get both regular and inverted oscillators.

It is worth stressing that the above Hamiltonian originates from the discretization of the
scalar field (\ref{eq:field}). This imposes a UV cut-off inverse proportional to the lattice spacing. As long as only the low energy physics is concerned, Hamiltonian (\ref{Ham2}) for the uncoupled oscillators should describe the original field theory very well.

Now we will talk about the structure of the quantum circuit we will be using to study the complexity. 
At $t=0$ we start with the ground state of $\tilde H (m, \Omega, \hat \lambda=0)$ and then time evolve it with $\tilde H (m, \Omega, \hat \lambda \neq 0)$ and $\tilde H' (m, \Omega, \hat \lambda' \neq 0)$ with two slightly different couplings, $\hat \lambda$ and $\hat \lambda'=\hat \lambda+\delta \hat \lambda,$ where $\delta \hat\lambda$ is small. The complexity of the state evolved by $H$ with respect to the state evolved by $H'$ is given by \cite{me3,Camargo:2018eof,Hackl:2018ptj}

\begin{equation}\label{eq:complexityN}
\mathcal{\hat C}(\tilde U) =\frac{1}{2}\sqrt{\sum_{k=0}^{N-1}\left(\cosh^{-1}\left[\frac{\omega_{r,k}^2+|\hat \omega_k(t)|^2}{2\,\omega_{r,k}\,\text {Re}  (\hat \omega_k(t))}\right]\right)^2},
\end{equation}
where
 \begin{align}
 \begin{split} 
 \hat \omega_k(t) =  i \ \Omega_k' \cot (\Omega_k' t)  + \frac{\Omega_k'^2}{  \sin^2( \Omega_k' t )\left(\omega_k(t) + i\, \Omega_k' \cot (\Omega_k' t)\right)}\end{split}
 \end{align}
 and
 \begin{align}
 \Omega_k'^2= \hat m^2+4\,(\Omega^2-\hat \lambda-\delta\hat \lambda)\sin^2\Big(\frac{\pi\,k}{N}\Big).
 \end{align}
 The frequencies-squared $\omega_k(t)^2,\ \omega_{r,k}^2$ are given by 
\begin{equation}
\begin{split} 
\label{def1}
\omega_k(t)=\Omega_k \left (\frac{\Omega_k-i\,\omega_{r,k} \cot (\Omega_k\, t)}{\omega_{r,k}-i\,\Omega_k \cot (\Omega_k\, t)}\right),
\end{split}
\end{equation}

Interestingly, this simple model exhibits three universal behaviors for the complexity growth in three different time scales. 

As shown in Fig.~\ref{fig_compgrowth}, the complexity starts to grow as a power-law, in a transient time known as the dissipation time \cite{Maldacena2016-mb}. This is a time scale when local perturbation relaxes. It corresponds to an exponential decay of time-ordered correlators of local observable. At larger times the complexity switches to an exponential growth, i.e., scrambling, which corresponds to the early exponential decay of the OTOC, $1-\epsilon e^{\lambda t}$. Asymptotically, the complexity grows linearly in time. This corresponds to the exponential relaxing of the OTOC. Note that the inverted harmonic oscillators are not bounded, the complexity grows forever without saturation.

We also identified the scaling of the complexity growth in terms of parameters $N$ and $\delta$ in the model. In the dissipation and intermediate linear growth regime, the complexity scales as $C\sim \sqrt{N}\delta^{-2}t$. In the scrambling regime the complexity grows as $C\sim \sqrt{N}\exp{\delta^{-2}t}$. The scrambling time, i.e., the time scale for which the complexity becomes $O(1)$, can then be extracted as $t_d \sim \delta^2\log{1/\sqrt{N}}$. As before, the conclusion drawn here is not sensitive to the choice of the cost functional (\ref{eq:complexityN}). This generic features of the complexity for this system still persists even if we use different cost functional.

\section{Discussion}
The harmonic oscillator is one of the most versatile toy models in all of physics. Largely, this is because the oscillator Hamiltonian is quadratic, and Gaussian integrals are a staple of any physicist's diet. This article details our systematic study of the {\it inverted} harmonic oscillator as a vehicle to explore quantum chaos and scrambling in a controlled and tractable setting. In particular, since the inverted harmonic oscillator is classically unstable but not chaotic, our expectation for the quantum system is to find scrambling, but not true chaotic behavior. We set out to ask if, and how, this expectation manifests at the level of some frequently used diagnotics. Concretely, we focused on two recently developed probes of chaos; the out-of-time-order correlator and the circuit complexity, both of which we computed for the displacement operator in eq.\eqref{eq:Doperator}. The OTOC in particular appears to be insensitive to whether or not the oscillator is inverted. On the other hand, the fact the oscillator Hamitonian is quadratic is a key feature of this computation. To test this, we extended the Hamiltonian by adding a cubic-gate perturbation and found that, with this additional term in the IHO Hamiltonian, the OTOC exhibits a crossover from no-decay to exponential-decay, consistent with the above expectation. We further computed the full quantum Lyapunov spectrum for the IHO, finding that it exhibits a paired structure among the Lyapunov exponents. This in turn leads us to conjecture that as long as the OTOC scrambles exponentially, such a structure will manifest in the Lyapunov spectrum. \\



 Using the operator method we then computed the complexity of a target displacement operator obtained from a simple reference displacement operator by the chaotic Gaussian dynamical evolution expected of the IHO. Our construction is primarily based on Nielsen's geometric formalism, making use of the Heisenberg group as a natural avatar for the displacement operator. The takeaway from this analysis is that the choice of operator or wave function approaches in the computation of the complexity really depends on the physical problem in question. For example, the wave function approach is more convenient for the study of a system of $N$-oscillators,  where we showed that both the complexity and scrambling time depend on two new parameters, namely, the number of oscillators and lattice spacing between them. \\

As elucidating as this study of the chaos and complexity properties of the inverted harmonic oscillator is, there are several questions remain to be addressed. Among these, we count:
\begin{itemize}
\item A clear recipe for the penalization procedure, which usually accompanies the circuit complexity, is still missing for continuous systems such as the inverted oscillator and, more pressingly, in quantum field theory. 
\item While progress in understanding circuit complexity has come in leaps and bounds since it entered into the horizon of high energy theory and black hole physics, much of this progress has been focused on simple linear systems. For the purposed of understanding {\it interacting} systems, it would be of obvious benefit to push the operator complexity computation beyond the Heisenberg group. Here too, the inverted oscillator offers some hints. For example, one conceptually straightforward extension that may shed some light into this matter would be precisely the cubic-gate perturbation that we considered above.   
\item Finally, and more speculatively, there is the novel and largely unexplored class of Hamiltonians with unbroken $\texttt{PT}$ symmetry which describe non-isolated systems in which the loss to, and gain from the environment are exactly balanced. In a sense then, \texttt{PT}-symmetric Hamiltonians interpolate between Hermitian and non-Hermitian Hamiltonians  but with spectra that are {\it real, positive and discrete}. An example relevant to our study here is the 1-parameter family of Hamiltonians
\begin{eqnarray*}
  H(\varepsilon) = \frac{p^{2}}{2} + x^{2}(ix)^{\varepsilon}\,,
\end{eqnarray*}
with real parameter $\varepsilon$. Clearly $H(0) = p^{2}/2 + x^{2}$ is just the familiar harmonic oscillator. On the other hand $H(1) = p^{2}/2 + ix^{3}$ is not only unfamiliar, it is also {\it complex}! Continuing along $\varepsilon$, we find the {\it inverted} quartic Hamiltonian $H(2) = p^{2}/2 - x^{4}$ which looks decidedly unstable. Nevertheless, it was rigorously shown in \cite{Dorey:2001uw} that the eigenvalues of $H(\varepsilon)$ are real for all $\varepsilon\geq0$. Given the numerous manifestations of \texttt{PT}-symmetric Hamiltonians in for example optics, superconductivity and even graphene systems, it would be interesting to explore both its quantum chaotic as well complexity properties with some of the tools that we have explored here.
\end{itemize}
We leave these and related questions for further study.

\section*{Note added in proof} After our article appeared on the arXiv, another article \cite{Hashimoto:2020xfr} was posted, discussing and further motivating the study of the inverted harmonic oscillator. There, using a recently developed technique for computing the thermal OTOC in single-particle quantum mechanics \cite{Hashimoto:2017oit}, the authors argue that the inverted harmonic oscillator emerges quite generically anytime one isolates one degree of freedom in a large N system with a gravity dual, and integrates out the remaining degrees of freedom - essentially forming a quadratic hilltop potential. They find an exponential growth of the OTOC with quantum Lyapunov exponent of order the classical Lyapunov exponent generated at the hilltop. In fact, they find that $\lambda_{\mathrm{OTOC}} \leq cT $ at temperature $T$ and for some constant $c\sim\mathcal{O}(1)$, therby generalizing the MSS chaos bound to single-particle quantum mechanics. This may seem to be in conflict with our finding for the OTOC, however, as pointed out above, we compute the OTOC for displacement operators which, being built out of ladder operators are {\it composite}. In this sense, our results are reminiscent of the operator thermalization hypothesis introduced in \cite{Sabella-Garnier:2019tsi}. The reconciliation of these findings is, no doubt, of great interest and we leave this for future work.
\section*{Acknowledgements}
A.B. is supported by Research Initiation Grant (RIG/0300) provided by IIT-Gandhinagar and Start Up Research Grant (SRG/2020/001380) by Department of Science \& Technology Science and Engineering Research Board (India). S.H. would like to thank the URC of the University of Cape Town for a research development grant for emerging researchers. J.M. was supported in part by the NRF of South Africa under grant CSUR 114599. W.A.C acknowledges support provided by the Institute for Quantum Information and Matter, Cal- tech, and for the stimulating environment from which the author had been significantly benefited. W.A.C gratefully acknowledges the support of the Natural Sciences and Engineering Research Council of Canada (NSERC). B.Y. was supported by the U.S. Department of Energy, Office of Science, Basic Energy Sciences, Materials Sciences and Engineering Division, Condensed Matter Theory Program. B.Y. also acknowledges partial support from the Center for Nonlinear Studies at LANL.



\bibliography{references.bib}

\end{document}